\documentclass[12pt]{article}
 \usepackage{epsfig}
 \def\be{\begin{equation}}
 \def\ee{\end{equation}}
 \def\bea{\begin{eqnarray}}
 \def\eea{\end{eqnarray}}
 \usepackage{graphicx}

 \catcode`\@=11
 \def\lsim{\mathrel{\mathpalette\@versim<}}
 \def\gsim{\mathrel{\mathpalette\@versim>}}
 \def\@versim#1#2{\vcenter{\offinterlineskip
 \ialign{$\m@th#1\hfil##\hfil$\crcr#2\crcr\sim\crcr } }}
 \catcode`\@=12

 \catcode`@=12
\begin{document}
 \thispagestyle{empty}
 \begin{flushright}
 UCRHEP-T610\\
 Feb 2021\
 \end{flushright}
 \vspace{0.6in}
 \begin{center}
 {\LARGE \bf Cobimaximal Mixing with Dirac Neutrinos\\}
 \vspace{1.2in}
 {\bf Ernest Ma\\}
 \vspace{0.2in}
{\sl Physics and Astronomy Department,\\ 
University of California, Riverside, California 92521, USA\\}
\end{center}
 \vspace{1.2in}

\begin{abstract}\
If neutrinos are Dirac, the conditions for cobimaximal mixing, i.e. 
$\theta_{23}=\pi/4$ and $\delta_{CP} = \pm \pi/2$ in the $3 \times 3$ 
neutrino mixing matrix, are derived.  One example with $A_4$ symmetry 
and radiative Dirac neutrino masses is presented.
\end{abstract}

 \newpage
 \baselineskip 24pt
\noindent \underline{\it Introduction}~:~
Neutrinos are mostly assumed to be Majorana.  The associated $3 \times 3$ 
mass matrix has been studied in numerous papers.  One particularly 
interesting form was discovered in 2002~\cite{bmv03}, i.e.
\begin{equation}
{\cal M}_\nu = \pmatrix{A & C & C^* \cr C & B & D \cr C^* & D & B^*},
\end{equation}
where $A$ and $D$ are real, which was shown subsequently~\cite{gl04} to be 
the result of a generalized $CP$ transformation involving $\nu_{\mu,\tau}$ 
exchange.  This form predicts the so-called cobimaximal mixing 
pattern~\cite{m16} of neutrinos, i.e. $\theta_{23} = \pi/4$ and 
$\delta_{CP} = \pm \pi/2$, which is close to what is observed~\cite{pdg20}.

To understand the cobimaximal mixing matrix $U_{CBM}$, consider its form  
in the PDG convention, i.e.
\begin{equation}
U_{CBM} = \pmatrix{c_{12} c_{13} & s_{12}c_{13} & \pm i s_{13} \cr 
-(1/\sqrt{2})(s_{12} \pm i c_{12} s_{13}) & (1/\sqrt{2})(c_{12} \mp i s_{12} 
s_{13}) & -c_{13}/\sqrt{2} \cr (1/\sqrt{2})(-s_{12} \pm i c_{12} s_{13}) & 
(1/\sqrt{2})(c_{12} \pm i s_{12} s_{13}) & c_{13}/\sqrt{2}}.
\end{equation}
Note that the $(2i)$ and $(3i)$ entries for $i=1,2,3$ are equal in magnitudes. 
It is easy then to obtain Eq.~(1) as
\begin{equation}
{\cal M}_\nu = U_{CBM} {\cal M}_{diag} U^T_{CBM}.
\end{equation}

In Eq.~(1), the neutrino basis is chosen for which the charged-lepton mass 
matrix ${\cal M}_l$ is diagonal which links the left-handed $(e,\mu,\tau)$ 
to their right-handed counterparts.  Suppose it is not, but rather that it 
is digonalized on the left by the special matrix~\cite{c78,w78}
\begin{equation}
U_\omega = {1 \over \sqrt{3}} \pmatrix{1 & 1 & 1 \cr 1 & \omega & \omega^2 
\cr 1 & \omega^2 & \omega},
\end{equation}
where $\omega = \exp(2\pi i/3) = -1/2 + i \sqrt{3}/2$.  It was 
discovered in 2000~\cite{fmty00} that 
\begin{equation}
U_{CBM} = U_\omega^\dagger {\cal O},
\end{equation}
where ${\cal O}$ is an orthogonal matrix.  The proof is very simple because 
the product $U_\omega {\cal O}$ enforces the equal magnitudes of the 
$(2i)$ and $(3i)$ entries.

The implications of these conditions regarding Dirac (instead of Majorana) 
neutrinos are the subject of this paper.  A specific model of radiative 
Dirac neutrino masses with $A_4$ symmetry~\cite{mr01} is also presented.

\noindent \underline{\it Form of Dirac Neutrino Mass Matrix}~:~
Consider the $3 \times 3$ mass matrix linking $\nu_L$ to $\nu_R$, i.e. 
${\cal M}_D$.  It is diagonalized by two unitary matrices, $U_L$ on the 
left and $U_R^\dagger$ on the right.  To eliminate $U_R$ which is 
unobservable in the standard model (SM) of quarks and leptons, the 
product ${\cal M}_D {\cal M}_D^\dagger$ should be studied.  It is 
automatically Hermitian, and is diagonalized by $U_L$ on the left and 
$U_L^\dagger$ on the right.  Using $U_{CBM}$ of Eq.~(2), it is easily 
seen that
\begin{equation}
{\cal M}_D {\cal M}_D^\dagger = \pmatrix{A & C & C^* \cr C^* & B & D \cr 
C & D^* & B}.
\end{equation}
This is the analog of Eq.~(1) for Dirac neutrinos.  An example was recently 
shown in Ref.~\cite{m21}.  However, Eq.~(6) does not constrain ${\cal M}_D$ 
uniquely because of the missing arbitrary $U_R$.

Nevertheless, a possible form of ${\cal M}_D$ is 
\begin{equation}
{\cal M}_D = \pmatrix{a & c & c^* \cr d & b & e \cr d^* & e^* & b^*},
\end{equation}
where $a$ is real.  It is then trivial to see that 
${\cal M}_D {\cal M}_D^\dagger$ yields exactly Eq.~(6).  The origin of 
this ${\cal M}_D$ is a simple extension of the generalized $CP$ transformation 
of Ref.~\cite{gl04}, i.e.
\begin{equation}
\nu_e \leftrightarrow \nu_e, ~~~ \nu_\mu \leftrightarrow \nu_\tau, ~~~ 
\nu_e^c \leftrightarrow \nu_e^c, ~~~ \nu_\mu^c \leftrightarrow \nu_\tau^c,
\end{equation}
together with complex conjugation.  It is important to realize that whereas 
Eq.~(7) guarantees Eq.~(6), the former may be obtained without the latter, 
as shown already in Ref.~\cite{m21} because of the missing arbitrary $U_R$.

\noindent 
\underline{\it Scotogenic Dirac  Neutrinos with Cobimaximal Mixing}~:~
The other approach to obtaining $U_{CBM}$ is through Eq.~(5).  Two  
previous models were constructed~\cite{m15,m16-1} this way for Majorana 
neutrinos.  Their Dirac counterpart is presented here.  It is actually 
simpler because a technical problem is naturally avoided in this case 
as shown below.

Following Ref.~\cite{mr01}, the non-Abelian discrete symmetry $A_4$ is 
used, under which the three families of left-handed lepton doublets 
transform as the $\underline{3}$ representation, and the three charged-lepton 
singlets as $\underline{1}, \underline{1}', \underline{1}''$.  There are 
also three Higgs doublets $\Phi_i = (\phi_i^+,\phi_i^0)$ transforming as 
$\underline{3}$.  The multiplication rules for two triplets $a_{1,2,3}$ 
and $b_{1,2,3}$ in this representation~\cite{mr01} are
\begin{equation}
a_1 b_1 + a_2 b_2 + a_3 b_3 = \underline{1}, ~~~ 
a_1 b_1 + \omega a_2 b_2 + \omega^2 a_3 b_3 = \underline{1}', ~~~ 
a_1 b_1 + \omega^2 a_2 b_2 + \omega a_3 b_3 = \underline{1}''.
\end{equation}
Assuming that $\langle \phi_i^0 \rangle$ is the same for $i=1,2,3$, the 
$3 \times 3$ mass matrix linking $(e,\mu,\tau)_L$ to $(e,\mu,\tau)_R$ is 
then
\begin{equation}
{\cal M}_l = U_\omega \pmatrix{m_e & 0 & 0 \cr 0 & m_\mu & 0 \cr 0 & 0 
& m_\tau},
\end{equation}
which is well-known since 2001.  

To obtain Dirac neutrinos, three lepton singlets $\nu_R$ transforming as 
$\underline{3}$ under $A_4$ are added to the SM.  Since 
\begin{equation}
\underline{3} \times \underline{3} = \underline{1} + \underline{1}' 
+ \underline{1}'' + \underline{3} + \underline{3},
\end{equation}
the products $(a_1 b_2 c_3 + a_2 b_3 c_1 + a_3 b_1 c_2)$ and  
$(a_1 b_3 c_2 + a_2 b_1 c_3 + a_3 b_2 c_1)$ are allowed, so that 
tree-level Dirac neutrino masses are obtained.  To forbid this, a 
$Z'_2$ symmetry is imposed, so that $\nu_R$ are odd, and the SM  
fields are even as shown in Table~1.
\begin{table}[tbh]
\centering
\begin{tabular}{|c|c|c|c|c|c|c|}
\hline
fermion/scalar & $SU(2)_L$ & $U(1)_Y$ & $A_4$ & $Z'_2$ & $Z_2^D$ & $L$ \\
\hline
$(\nu,l)_{iL}$ & 2 & $-1/2$ & 3 & $+$ & + & 1 \\ 
$l_{iR}$ & 1 & $-1$ & $1,1',1''$ & $+$ & + & 1 \\
$\nu_{iR}$ & 1 & 0 & $3$ & $-$ & + & 1 \\ 
\hline
$(\phi^+,\phi^0)_i$ & 2 & 1/2 & 3 & $+$ & + & 0 \\ 
$(\eta^+,\eta^0)$ & 2 & 1/2 & 1 & $+$ & + & 0 \\ 
$s_i$ & 1 & 0 & 3 & $+$ & $-$ & 0 \\ 
$s'_i$ & 1 & 0 & 3 & $-$ & $-$ & 0 \\
\hline
$(E^0,E^-)_{L,R}$ & 2 & $-1/2$ & 1 & + & $-$ & 1 \\ 
$N_{L,R}$ & 1 & $0$ & 1 & + & $-$ & 1 \\ 
\hline
\end{tabular}
\caption{Fermion and scalar content of Dirac neutrino model with dark 
$Z_2^D$ and $A_4 \times Z'_2$ symmetries.}
\end{table}
Note that all dimension-four terms in the Lagrangian are required to obey 
$A_4 \times Z'_2$ which is only broken together softly by the $ss'$ mass 
terms.   Added are dark scalars and fermions which are odd under an exactly 
conserved $Z_2^D$ symmetry.  Lepton number $L$ is conserved as shown.  Dirac 
neutrino masses are radiatively generated by dark matter~\cite{fm12} as shown 
in Fig.~1 in analogy with the original scotogenic model~\cite{m06}.  
The key for cobimaximal mixing is that the $s,s'$ scalars are real 
fields~\cite{m15,m16-1}. 
\begin{figure}[htb]
\vspace*{-5cm}
\hspace*{-3cm}
\includegraphics[scale=1.0]{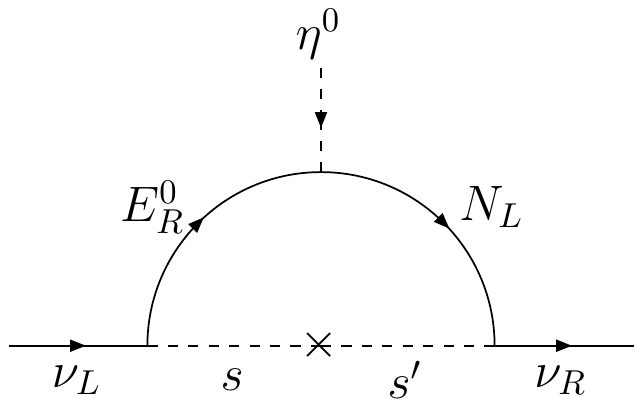}
\vspace*{-21.5cm}
\caption{Scotogenic Dirac neutrino mass.}
\end{figure}

The relevant Yukawa couplings are
\begin{equation}
f_E s \bar{E}_R^0 \nu_L, ~~~ f_N s' \bar{\nu}_R N_L, ~~~ 
f_{NE} \eta^0 \bar{N}_L E_R^0, ~~~ f_{EN} \bar{\eta}^0 \bar{E}_L^0 N_R.
\end{equation}
All respect $A_4 \times Z'_2$, with the latter two contributing to the 
$2 \times 2$ mass matrix linking $(E^0_L,N_L)$ to $(E^0_R,N_R)$, i.e.
\begin{equation}
{\cal M}_{EN} = \pmatrix{m_E & m_{EN} \cr m_{NE} & m_N} = 
\pmatrix{\cos \theta_L & -\sin \theta_L \cr \sin \theta_L & \cos \theta_L} 
\pmatrix{m_1 & 0 \cr 0 & m_2} 
\pmatrix{\cos \theta_R & \sin \theta_R \cr \sin \theta_R & \cos \theta_R}.
\end{equation}
As for the contribution of $s$ and $s'$, the mass-squared matrix for each 
is proportional to the identity, whereas the $s s'$ mixing is arbitrary, 
breaking both $A_4$ and $Z'_2$ at the same time softly.  Let it be denoted 
as ${\cal M}^2_{ss'}$ and assuming that its entries are all much smaller 
then the invariant masses of $s$ and $s'$, then it is clear that the 
Dirac neutrino mass matrix in the basis of Fig.~1 is proportional to 
${\cal M}^2_{ss'}$ and is real up to an 
unobservable phase, i.e. the relative phase of $f_N$ and $f_E$.  This 
means that it is diagonalized by an orthogonal matrix.  Combined with 
Eq.~(10), cobimaximal mixing is assured.

The explicit expression for the scotogenic Dirac neutrino mass matrix is 
\begin{equation}
{\cal M}_\nu = {f_N^* f_E {\cal M}^2_{ss'} \over 16 \pi^2} \left[ 
{\sin \theta_L \cos \theta_R \over m_1} {[F(x_1)-F(y_1)] \over [x_1-y_1]} - 
{\cos \theta_L \sin \theta_R \over m_2} {[F(x_2)-F(y_2)] \over [x_2-y_2]} 
\right],
\end{equation}
where $x_{1,2} = m^2_s/m_{1,2}^2$ and $y_{1,2} = m^2_{s'}/m_{1,2}^2$, 
and
\begin{equation} 
F(x) = {x \ln x \over x-1}.
\end{equation}

\noindent \underline{\it The $A_4 \to Z_3$ Breaking}~:~
The breaking of $A_4$ by $\langle \phi_i^0 \rangle = v$ reduces this 
symmetry to $Z_3$~\cite{m10}.  It must be maintained for $U_\omega$ to be 
valid. However, the addition of $s$ and $s'$ would allow the quartic terms 
$s_i s_j \Phi_i^\dagger \Phi_j$ and $s'_i s'_j \Phi_i^\dagger \Phi_j$. 
The key now is that both the quadratic mass terms $s_i s_i$ and $s'_i s'_i$
do not break $Z'_2$ and are required also not to break $A_4$.  Only the 
$s_i s'_j$ terms break both $A_4$ and $Z'_2$ softly together.  Hence the 
one-loop correction to $\Phi_i^\dagger \Phi_j$ is shown in Fig.~2. 
\begin{figure}[htb]
\vspace*{-5cm}
\hspace*{-3cm}
\includegraphics[scale=1.0]{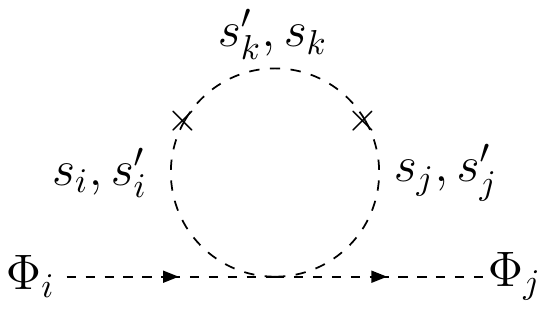}
\vspace*{-21.5cm}
\caption{Finite one-loop correction to $\Phi_j^\dagger \Phi_j$.}
\end{figure}
Two mass insertions are required, which render the diagram finite and 
suppressed so that the residual $Z_3$ symmetry is maintained to a good 
approximation.  In Refs.~\cite{m15,m16-1}, this option is not available 
for Majorana neutrinos because $s'$ is absent and $s_i s_j$ breaks $A_4$, 
which yields only one mass insertion in Fig.~2, thus making it 
logarithmically divergent.  

\noindent \underline{\it Dark Sector}~:~
The dark sector fermions are $(E^0,E^-)$ and $N$.  The two neutral ones 
have masses $m_{1,2}$ and the charged one $m_E$.  They are assumed greater 
than the masses of the scalars, $m_s$ and $m_{s'}$, with the small 
${\cal M}^2_{ss'}$ mixing between them.  Let $m_s$ be the smaller, 
then the almost degenerate $s_{1,2,3}$ are dark-matter candidates. 
They interact with the SM Higgs boson $h$ according to
\begin{equation}
{\cal L}_{int} = -\lambda s_i s_i \left( vh + {h^2 \over 2} \right)  
-\lambda' s_i s_j \left( vh + {h^2 \over 2} \right),
\end{equation}
where
\begin{equation}
h = {\sqrt{2} [\langle \eta^0 \rangle Re(\eta^0) + \langle \phi_i^0 \rangle 
Re(\phi_1^0 + \phi_2^0 + \phi_3^0)] \over \sqrt{\langle \eta^0\rangle^2 + 
3 \langle \phi_i^0 \rangle^2}}.
\end{equation}
This is a straightforward generation of the simplest model of dark matter, 
i.e. that of a real scalar.  The comprehensive analysis of 
Ref.~\cite{gambit17} is thus applicable.

\noindent \underline{\it Conclusion}~:~
It is shown how cobimaximal neutrino mixing, i.e. $\theta_{23}=\pi/4$ and 
$\delta_{CP}=\pm\pi/2$, occurs for Dirac neutrinos.  It is defined by 
Eq.~(6) which is obtainable from Eq.~(7) based on a generalized $CP$ 
transformation.  However, because of the missing arbitrary unitary matrix 
$U_R$ which diagonalizes ${\cal M}_D$ on the right, there are certainly 
other solutions, one of which is discussed in Ref.~\cite{m21}. 

Another approach is to use Eq.~(5), which may be implemented with the 
non-Abelian discrete symmetry $A_4$ and a scotogenic Dirac neutrino 
mass matrix proportional to a real scalar mass-squared matrix.  It is 
the analog of previous suggestions~\cite{m15,m16-1} for Majorana neutrinos, 
but in the case of Dirac neutrinos here, it is more technically natural.

\noindent \underline{\it Acknowledgement}~:~
This work was supported in part by the U.~S.~Department of Energy Grant 
No. DE-SC0008541.

\bibliographystyle{unsrt}

\end{document}